\documentclass[10pt,letterpaper,twocolumn]{article} 

\usepackage{ol2}
\usepackage[draft,implicit=false]{hyperref}
\usepackage{amsmath}

\begin{document}

\twocolumn[ 

\title{Exact cascading nonlinearity in quasi-phase-matched quadratic media}


\author{Matteo Conforti}

\address{
Dipartimento di Ingegneria dell'Informazione, Universit\`a di Brescia, Via Branze 38, 25123 Brescia, Italy\\
PhLAM/IRCICA CNRS-Universit\'e Lille 1, UMR 8523/USR 3380, F-59655 Villeneuve d'Ascq, France\\
email: confortimatteo@gmail.com
}

\begin{abstract}
The evolution of light pulses and beams in a quasi-phase-matched (QPM) quadratic medium is usually described by considering only the spatial harmonic of the QPM grating that minimizes the residual phase-mismatch. I show that, for strongly phase-mismatched interactions (the cascading regime), several harmonics need to be accounted for in order to obtain the correct value of the effective cubic nonlinearity, of which I find a simple analytical expression. I discuss the effects of the higher order harmonics of the grating on solitary wave propagation.
\end{abstract}


 ] 

\noindent 
Second harmonic generation (SHG) was the first observed nonlinear optical phenomenon, and it has been extensively exploited for frequency conversion.
When the wave-number mismatch between the fundamental frequency (FF) and the second harmonic (SH) fields is high, consecutive energy exchanges between FF and SH take place, giving rise to the phenomenon of cascading \cite{desalvo92}, that produces an effective cubic nonlinearity. The magnitude and sing of the cascading nonlinearity is controlled by the phase-mismatch and can sustain the propagation  of spatial or temporal solitons \cite{menyuk94,bache07}, even in the normal dispersion regime. Cascading is still attracting much interest, in particular for time domain application, as pulse compression \cite{zhou12} and ultrabroadband spectrum generation \cite{levenius12}.

In the context of second-harmonic generation, quasi-phase-matching (QPM) is a very well-known technique, relying on the periodic modulation of the nonlinear susceptibility, that can compensate for the phase-mismatch between the FF and SH wave-number, through and additional grating wave-vector.
Moreover, QPM structures can be engineered by varying the period or duty cycle of the grating, enabling several applications in the context of pulse shaping \cite{reid05,baronio06,reid08,conforti07,marangoni09,fejer13}, and supercontinuum generation \cite{fejer07,fejer11,conforti10}.
It has been shown that near first-order QPM (provided by the fundamental harmonic of the grating), the high order harmonics induce cubic nonlinear self-phase-modulation and cross-phase-modulation terms in the equations for the averaged fields \cite{clausen97}. These high-order effects, that are unavoidable in physically realizable gratings (typically square waves),  can support peculiar spatial solitons \cite{clausen97}, can be exploited for all-optical switching \cite{kobyakov98}, and can be engineered to overwhelm the intrinsic cubic nonlinearity of the medium \cite{bang99}. The fundamental assumptions of these studies is that the grating is designed to be very close to first order QPM.

In this letter I show that the high order harmonics of the grating must be accounted for also in the cascading regime. I find a simple analytical expression for the induced self phase modulation coefficient, that may differ drastically from the conventionally adopted value $-(\chi_{eff})^2/\delta k$ \cite{baronio06}. Moreover, the difference is bigger when the cascading approximation is more accurate (i.e. for high residual phase-mismatches). I discuss the effects on propagation of temporal solitons in Lithium Niobate for poling periods, pulse durations and wavelengths typically encountered in applications.

The coupled equations ruling the evolution of the FF and SH electric field envelopes are:
\begin{eqnarray}
i\frac{\partial A_1}{\partial z}-\frac{\beta_1''}{2}\frac{\partial^2 A_1}{\partial t^2}+g(z)\chi A_1^*A_2e^{i\Delta k z}=0,\label{FF}\\
i\frac{\partial A_2}{\partial z}+i\delta\frac{\partial A_2}{\partial t}-\frac{\beta_2''}{2}\frac{\partial^2 A_2}{\partial t^2}+g(z)\chi A_1^2e^{-i\Delta k z}=0, \label{SH}
\end{eqnarray}
where $\beta_{1,2}''$ are the group velocity dispersion (GVD) at FF and SH, $\chi=\omega_0d_{33}/(n_0c)$ is the effective quadratic nonlinear coefficient,
$\Delta k=k_2-2k_1$ is the natural phase mismatch, $k_1=k(\omega_0)$, $k_2=k(2\omega_0)$, $\delta$ is the group velocity mismatch (GVM) between FF and SH, and 

\begin{equation}
g(z)=\sum_{n=-\infty}^{+\infty}\frac{2}{i\pi(2n+1)}e^{i(2n+1)\kappa z},\;\kappa=\frac{2\pi}{\Lambda}
\end{equation}

is the Fourier expansion of the QPM grating of period $\Lambda$.
The same equations rule the propagation of beams, making the substitutions $t\rightarrow x$, $\beta_{1,2}''\rightarrow -1/k_{1,2}$, $\delta \rightarrow \alpha$, being $\alpha$ the spatial walk-off.

In the cascading limit, FF can be considered as nearly constant, and the equation (\ref{SH}) can be integrated to yield, as first order approximation:
\begin{equation}\label{A2}
A_2\approx\frac{2\chi}{i\pi}A_1^2\sum_{m=odd}\frac{e^{i(\kappa m-\Delta k)z}-1}{m(\kappa m -\Delta k)}.
\end{equation}

By substituting back the expression (\ref{A2}) into Eq. (\ref{FF}), the nonlinear term NL of the resulting equation reads as:
\begin{equation}
NL=-\left(\frac{2\chi}{\pi}\right)^2|A_1|^2A_1\sum_{m,n=odd}\frac{e^{i(m+n)\kappa z}-e^{i(m\kappa+\Delta k)z}}{mn(n\kappa-\Delta k)}.
\end{equation}

In the cascading regime, the residual phase mismatch must be high for all QPM orders ($\Delta k +\kappa m\gg1, \,\forall m$); moreover
the nonlinear interaction can be efficient only if $NL$ does not contain rapidly oscillating terms. This can be obtained only if one considers $m=-n$ in the double sum. In this case, Eq. (1) reduces to the nonlinear Shr\"odinger equation (NLS):
\begin{equation}
i\frac{\partial A_1}{\partial z}-\frac{\beta_1''}{2}\frac{\partial^2 A_1}{\partial t^2}+\gamma |A_1|^2A_1,\label{NLS}
\end{equation}

where the effective cubic coefficient $\gamma$ reads as:
\begin{equation}\label{gamma}
\gamma=-\left(\frac{2\chi}{\pi}\right)^2\sum_{n=odd}\frac{1}{n^2(\Delta k-n\kappa)}=-\left(\frac{2\chi}{\pi}\right)^2 S.
\end{equation}

The sum of the series in (\ref{gamma}) can be calculated, to get:
\begin{equation}\label{S}
S=\left\{
\begin{array}{cc}
\displaystyle \ \frac{\pi^2}{4\Delta k},~&\kappa = 0\\
 \displaystyle \ \frac{\pi^2}{4\Delta k}-\frac{\pi}{2}\frac{\kappa}{\Delta k^2}\tan\left(\frac{\pi}{2}\frac{\Delta k}{\kappa}\right),~&\kappa\neq 0 
\end{array} \right.
\end{equation}

In a homogeneous medium $\kappa=0$, and Eq. (\ref{gamma}) turns out to be

\begin{figure}[!tb]
\centerline{\includegraphics[width=7.5cm]{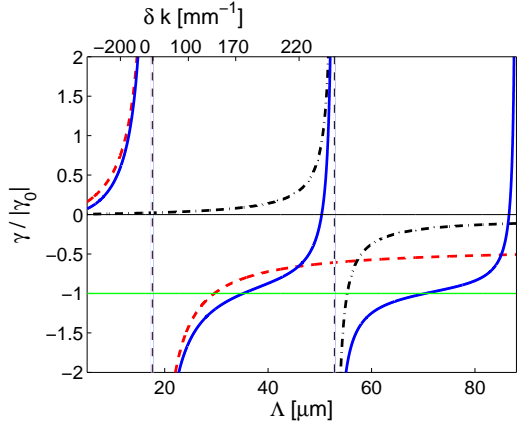}}
\caption{(Color online) Effective cascading nonlinearity $\gamma$ normalized to the homogeneous medium nonlinearity $\gamma_0$ (solid green line). Solid blue curve, exact value Eq. (\ref{gamma}); dashed red curve, standard first order QPM approximation [Eq. (\ref{gamma1}), $m=1$]; dash-dotted black curve, third-order QPM approximation [Eq. (\ref{gamma1}), $m=3$].  }\label{fig1}
\end{figure}

\begin{equation}\label{gamma0}
\gamma_{0}=-\frac{\chi^2}{\Delta k},
\end{equation}
as expected from the standard analysis \cite{menyuk94}. Interesingly enough, it has been demonstrated very recently that, despite the huge phase-mismatch, strong nonlinear interactions can take place in bulk media, and they can be exploited for pulse compression down to the single cycle regime \cite{zhou12}.

In a QPM medium, the series periodically diverges when $\Delta k=m\kappa$, i.e. for the $m-$th order quasi-phase matching. In these cases there is a strong conversion to SH and the cascading reduction breaks down.

Considering a single spatial harmonic  $m=2n+1$, we get:

\begin{equation}\label{gamma1}
\gamma_{m}=-\left(\frac{2\chi}{m\pi}\right)^2\frac{1}{\delta k},\, \delta k= \Delta k -m\frac{2\pi}{\Lambda}.
\end{equation}

Formula (\ref{gamma1}) is the well-known expression of $m-$th order QPM cascading nonlinearity. However, I show in the following that this estimate turns out to be very rough, and can be affected by significant error.


\begin{figure}[!b]
\centerline{\includegraphics[width=7.5cm]{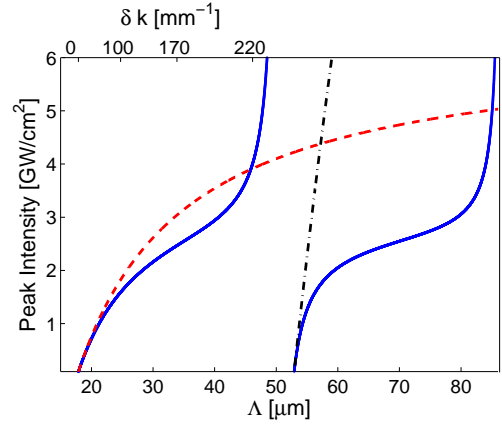}}
\caption{(Color online) Soliton peak intensity for a duration $T_0=40fs$ ($\lambda_0=1500nm$) as a function of QPM period. Solid blue curve, exact value; Dashed red curve, first order QPM approximation; Dash-dotted black curve, third order QPM approximation;}\label{fig2}
\end{figure}

Figure \ref{fig1} reports the exact value of the cascading nonlinearity $\gamma$ together with its approximations $\gamma_1$ and $\gamma_{3}$, as a function of QPM period $\Lambda$ and residual mismatch $\delta k=\Delta k-2\pi/\Lambda$ . It is considered the relevant example of a Lithium Niobate crystal, for a FF wavelength $\lambda_0=1500 nm$.
 It is clear that the standard first order QPM approximation holds only in a regime of moderate phase-mismatch $|\delta k|< 50mm^{-1}$. Generally speaking, considering only a single QPM harmonic works well only very close to phase-matching points, exactly where the cascading approximation is poor. 
The single harmonic approximation is affected by even more error for high order QPM (see dashed-dotted curve for $m=3$ QPM order).

\begin{figure}[!b]
\centerline{\includegraphics[width=7.5cm]{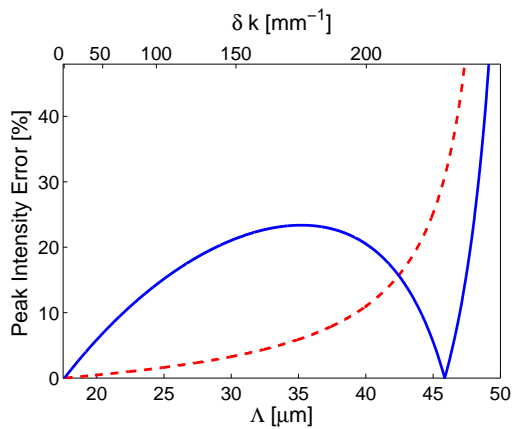}}
\caption{(Color online) Relative error for the soliton peak intensity: Solid blue curve, single-harmonic first order QPM approximation; dashed red curve two-harmonic first order QPM approximation.} \label{fig3}
\end{figure}

It is interesting to study the effects of QPM on cascading solitons of NLS (\ref{NLS}). In normal dispersion ($\beta_1>0$, typical of quadratic crystal in the visible and near-IR range), solitons exist if the cubic nonlinearity is defocusing ($\gamma<0$). Soliton peak intensity is proportional to $|\beta_1''/\gamma|/T_0^2$, and an example reported in Fig. \ref{fig2} considering a pulse duration $T_0=40fs$. Of course the errors on the cubic coefficients results in wrong estimates of the soliton peak intensity. For first order QPM grating (around $20\mu m<\Lambda <40\mu m$), error can be as big as $30\%$ (see Fig. \ref{fig3}), before diverging near phase-matching points. A more accurate approximation in this range can be obtained by considering both the positive and negative fundamental harmonics of the grating [$\gamma_{1,-1}=-2(2\chi/\pi)^2\Delta k/(\Delta k^2-\kappa^2)$, dotted curve in Fig. \ref{fig3}]. 
For longer periods (second order QPM) the estimates is totally wrong (dashed-dotted curve in Fig. \ref{fig2}), heavily overestimating the real peak intensity.

\begin{figure}[!tb]
\centerline{\includegraphics[width=7.5cm]{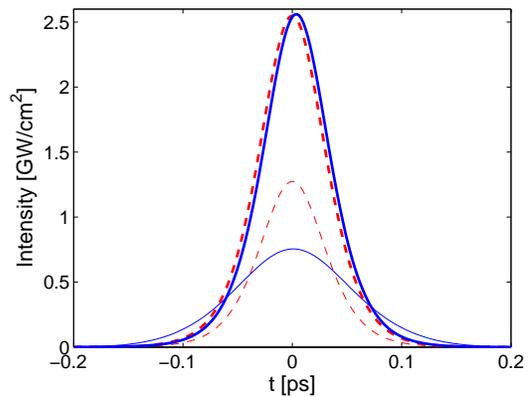}}
\caption{(Color online) Input (dashed curves) and output (solid curves) intensities for a sech pulse of duration $T_0=40fs$ and wavelength $\lambda=1500nm$ after propagation in a $4cm$ long crystal (around 3 diffraction lengths) poled with period $\Lambda=70\mu m$. Thick curves,  soliton case; thin curves, quasi-linear regime (half input intensity).} \label{fig4}
\end{figure}

To confirm the theoretical results, Eqs. (\ref{FF}-\ref{SH}) have been solved by standard split-step Fourier method, using as input condition the soliton profile $A_1(0,t)=\sqrt{|\beta_1''/\gamma|}/T_0sech(t/T_0)$, $A_2(0,t)=0$. For example, let us consider a $T_0=40fs$ pulse at $\lambda_0=1500nm$ propagating in a $4cm$-long Lithium Niobate crystals, poled with a period $\Lambda=70\mu m$. Figure \ref{fig4} shows input and output intensities (thick curves): after propagating for approximately three diffraction lengths, the pulse shape is nearly undistorted. This fact also witnesses the robustness of the solitons to secondary effects such as high-order dispersion, or GVM-induced self-steepening \cite{baronio06}. For comparison output intensity is reported for a pulse propagating in a quasi linear regime (half the input intensity needed to form the soliton).

It is worth to remind that quadratic media posses an intrinsic positive cubic nonlinearity. It simply acts as a bias, that adds itself to the total cubic coefficient. Moreover, for the wavelengths in the infrared (say $\lambda>1200nm$ for Lithium Niobate  \cite{zhou12}), it turns out to be much smaller than bulk cascading term $\gamma_0$.

In conclusion, I showed that high order harmonics of a QPM grating must be accounted for in the cascading regime. An simple analytical expression is reported for the induced self phase modulation coefficient, that may differ significantly from the conventionally adopted value. I discussed the effects on propagation of temporal solitons in Lithium Niobate for conventionally used poling periods, pulse duration and wavelengths.

Funding from MIUR (Grants No. PRIN 2009P3K72Z and No. PRIN 2012BFNWZ2) and ANR (grant TOPWAVE)
 is gratefully acknowledged.


\clearpage

\section*{Informational Fourth Page}

\end{document}